\documentstyle[aps,prl]{revtex}
 
 \def\beq{\begin{equation}}
 \def\eeq{\end{equation}}
 \def\beqa{\begin{eqnarray}}
 \def\eeqa{\end{eqnarray}}
 \def\lf{\nonumber \\}

 \gdef\s#1{\! #1 \!}
 \gdef\l#1{\> #1 \>}

\begin{document}
\twocolumn
\narrowtext

\noindent {\large \bf
On the absence of Shapiro-like steps in certain
mesoscopic {\it S-N-S} junctions }
\medskip

In DC transport through mesoscopic {\it S-N-S} junctions, it has been
observed \cite{Courtois} that as the temperature is increased the
Josephson coupling rapidly disappears, but the phase dependence
of the conductance persists to much higher temperatures and decreases 
only as $1/T$.  In a recent Letter \cite{Volkov} AC dynamical effects
in such junctions were discussed, and it was shown that this type
of phase dependent conductance should lead to Shapiro-step-like
features, which would also decrease in size only as $1/T$.   
In this Comment it is pointed out that such a phase-dependent
conductance cannot, by itself, lead to the formation of Shapiro steps.

A simple model for an {\it S-N-S} junction is given by the following
expression for the current $I$:
\beq \label{igv}
I  \l=  J(\phi) + G(\phi) V  \; ,
\eeq
where $J(\phi)$ is the Josephson current, $\phi$ is the
superconducting phase difference across the junction, $V = d\phi / dt$
is the voltage, and $G(\phi)$ is the shunt conductance ($I$, $\phi$
and $V$ are time dependent quantities; we have set $\hbar/2e=1$).  In
the standard resistively-shunted-junction (RSJ) model, the conductance
$G$ is taken to be independent of $\phi$.  When the junction is driven
by a DC and an AC bias, Eq.~(1) allows for phase locking between the
AC Josephson effect and the AC drive, over a range of values of the
DC current.  This gives rise to Shapiro steps in the time-averaged
{\it I-V} curve.

In long diffusive junctions such as those considered in 
Ref.~\cite{Volkov}, the Josephson current becomes vanishingly small,
$J(\phi) \s= 0$, at temperatures which are still well below the critical
temperature of the superconducting electrodes.
In this case no phase locking takes place, as can be seen by
directly integrating Eq.~(1) over time \cite{Maui}
\beqa \label{proof}
I_{DC}  & \l= &  \lim_{T \rightarrow \infty} 
{1 \over T} \int_0^T G(\phi(t)) \> {d\phi \over dt} \> dt
\lf & \l= &  \lim_{T \rightarrow \infty} 
{1 \over T} \int_{\phi(0)}^{\phi(T)} G(\phi) \> d\phi
\lf & \l= &
 \overline G \> V_{DC}  \; .
\eeqa
Here $\overline G = \int_0^{2\pi} G(\phi) \, d\phi /2\pi$ is the
phase-averaged conductance, and $I_{DC}$ and 
$V_{DC} = \lim_{\, T \rightarrow \infty \,} [\phi(T) \s- \phi(0)]/T$ 
are the time-averages, or DC components, of $I$ and $V$.
Thus, the time averaged {\it I-V} curve for a junction described by
Eq.~(1) with $J(\phi) \s= 0$ is completely unaffected by the
phase-dependent part of $G$ --- in such junctions an AC drive does not
bring about Shapiro steps.   This is independent of the nature of the
applied bias (current or voltage).

The structures considered in both the theoretical \cite{Volkov} and
the experimental \cite{Courtois,Harris} work are somewhat more
involved, with additional external normal leads.  A normal current
$I_N$ flows through these leads in response to a voltage $V_N$ applied
to them.  The corresponding DC differential conductance, $G_N$ was
shown in Ref.~\cite{Volkov} to be multivalued when the frequency of
the AC Josephson effect is commensurate with that of an AC drive.
This means that a plot of $G_N$ {\it vs.} $V_{DC}$ would exhibit
dramatic features (spikes and dips) at certain values of $V_{DC}$,
even in the absence of Josephson currents.  We point out that such an
effect would differ substantially from the standard Shapiro steps: in
junctions of the type considered here, $V_{DC}$ is always equal to
$I_{DC} / \, \overline G$, so no phase-locking mechanism is present.
In order to observe the effects considered in Ref.~\cite{Volkov}, one
would have to either precisely tune $V_{DC}$ and/or the AC drive to
make their frequencies equal and control their relative phase, or rely
on a phase-locking mechanism which has not yet been identified.  The
range in $V_{DC}$ (or $I_{DC}$) over which such phase locking would
occur will depend on the nature of this unidentified mechanism.

This work was supported by ONR, grant No.\ N00014-92-J-1452, by NSF, 
grant No.\ PHY94--07194, and by QUEST, a National Science Foundation
Science and Technology Center, (NSF grant No.\ DMR91--20007). 
N. A. acknowledges the support of a Fulbright fellowship.

\bigskip\medskip

{\small \noindent 
J.G.E. Harris$^*$, N. Argaman$^\dagger$, and S.J. Allen$^*$ \\
\indent\hspace*{1pt}University of California \\
\indent\hspace*{1pt}Santa Barbara, CA 93106

\bigskip

\noindent Revised September 9th, 1996

\bigskip\medskip

\indent $^*$ Physics Department.\\
\indent $^\dagger$ Institute for Theoretical Physics.

}

\end{document}